\def\sqig{$\sim\,$} \def\etal{et\,al.}  
\def\up#1{$^{\mbox{{\scriptsize #1}}}$} 
\def\deg{$^{\circ}$} \def\kmps{km\,s$^{-1}$} 
\def\Hb{H$\beta$} \def\HeI{He\,{\sc i}} \def\HeIIl{He\,{\sc ii}\ $\lambda$4686} 
\def\sqiglt{\hbox{\rlap{\lower.55ex \hbox {$\sim$}}
        \kern-.3em \raise.4ex \hbox{$<$}\,}}
\def\sqiggt{\hbox{\rlap{\lower.55ex \hbox {$\sim$}}
        \kern-.3em \raise.4ex \hbox{$>$}\,}}
\documentstyle[aas2pp4]{article}
\lefthead{Hellier}
\righthead{Intermediate polars}
\begin{document}
\title{Doppler tomography of intermediate polar spin cycles}
\author{Coel Hellier}
\affil{Department of Physics, Keele University, Keele, Staffordshire, ST5 5BG,
U.K.\\ Electronic mail: ch@astro.keele.ac.uk}
\begin{abstract}
I collect emission line profiles for seven intermediate polars,
phase-resolved over the spin cycles.  I then present the first
systematic application of Doppler tomography to the spin cycles,
with the aim of investigating 
whether spin-cycle tomography is a useful diagnostic tool.
Four stars show line emission from material falling onto both magnetic 
poles of the white dwarf, but three others show line emission from only 
the upper pole. Tomography can map the extent
and shape of the line-emitting accretion curtains, but it can also give
misleading results if the line flux changes significantly over the
spin cycle. 
\end{abstract}

\keywords{accretion, accretion disks --- binaries, close --- 
stars: magnetic fields --- novae, cataclysmic variables --- 
techniques: spectroscopic \\ [3cm] \hspace*{5cm} Accepted for ApJSupp 1999}

\section{Introduction}
Owing to the Doppler shift, an element of material in 
circular motion produces spectral lines with a sinusoidal modulation of
wavelength. Observations of such `S-waves' in time-resolved spectra
are a primary diagnostic for many fields. For instance, Vogt \&\ Penrod
(1983) developed a Doppler imaging technique to map line-profile
changes onto star spots on the rotating surfaces of active stars. 
A similar, Doppler tomography technique was first applied to close
binaries by Marsh \&\ Horne (1988). It inverts a phase-resolved
spectrum into a 2-dimensional velocity map in the rotating frame.
The amplitude and phase of each component S-wave of
a phase-resolved spectrum maps onto a particular x, y location 
on the velocity map.

The original implementation of Doppler tomography ensured a unique velocity 
map by using a maximum entropy method. However, a computationally-easier 
implementation, Fourier-filtered back-projection, has become 
more popular. This sums the flux along the track of each possible
S-wave to fill out the velocity map (see, e.g., Robinson, Marsh \&\ Smak
1993).

Tomography of the orbital cycle in cataclysmic variables (CVs) has become 
common, with too many instances now in the literature to list here, although 
the extensive compilation by Kaitchuck \etal\ (1994) is worth noting
(this paper, Marsh \&\ Horne 1998, and Robinson \etal\ 1993 are good 
tutorials for those unfamiliar with tomography applied to cataclysmic variables).
The technique works well for non-magnetic CVs, and for the 
AM Her stars, where the spin cycle of the magnetic white dwarf is locked
to the binary orbit (see Schwope, Mantel \&\ Horne 1997 for sample
tomograms). 

However, in magnetic CVs which aren't phase-locked
(the intermediate polars) the emission lines also vary with the
white-dwarf spin cycle as the magnetically-threaded material is wheeled 
round by the dipolar field. Variations on more than one periodicity 
aren't accounted for in standard tomography.
One remedy is to calculate tomograms on the orbital
and spin cycles separately, assuming that variations on the other cycle
smear into a phase-invariant profile that can be subtracted.  
The process of calculating a tomogram on a spin cycle is identical to
that of calculating an orbital tomogram --- the tomography algorithm
simply projects or de-projects between circular motion and trailed
spectra, without caring whether the motion is orbital or spin related.
The only difference is that, while orbital tomography should reveal
the structures fixed in the binary frame (the disc, secondary star
and accretion stream), spin-cycle tomography should instead provide 
velocity maps of the `accretion curtains' of magnetically channelled 
material falling onto the white dwarf.  

This technique has so far been used only twice, in the papers 
Hellier (1997) and Still, Duck \&\ Marsh (1998), but several other 
datasets suitable for tomography are in the literature. Here, I present 
a compilation of spin-cycle tomograms from old and new data, with the aim 
of allowing comparisons between different stars and of investigating 
whether spin-cycle tomography is a useful diagnostic tool.

A review of intermediate polars (and references for the stars discussed
here) is given by Patterson (1994). The standard picture supposes
that circulating material (conventionally a disk) feeds material onto
the magnetosphere (extending to \sqig 10 white dwarf radii), and 
that inside the magnetosphere the accreting material follows the field
lines of a dipolar field, which is tilted with respect to the spin axis. 
The tilt causes the field to pick up material preferentially from the
region of the disk to which the pole points, so that material flows
to the upper pole from one side of the disk and to the lower pole from the
other side (several papers,
including Hellier, Cropper \&\ Mason 1991, present illustrations of this
process). The azimuthal extents of the resulting accretion curtains 
are poorly known, with little guidance from theory.

\section{Data and Method}
I present spin-cycle tomograms of the emission lines of seven intermediate
polars (they are listed in Table~1 along with the paper in which the data
originally appeared). An eighth star, TV~Col, is not included because the
spin-cycle variations of the emission lines are too weak (Hellier 1993a).
For details of the data analysis and further discussion of the data see the
individual papers. 

For this paper the first step was to plot the line profile as a function of
the spin period (uppermost panels in each section of Fig.~1).  The
spin-varying component of the emission will come mostly from within the
rotating magnetosphere (although some could result from irradiation of an
accretion disk by spin-pulsed X-rays). Emission from further out (the disk, 
secondary star and stream) is expected to vary on the orbital cycle  
or on the beat cycle between the spin and orbital cycles. Thus
the spin-folded data will be diluted by these components. The second step,
therefore, was to subtract the phase-invariant profile (the lowest value at
each wavelength) to leave only the spin-varying component (shown in the
panels second from top in each section of Fig.~1). If the subtraction is
imperfect, the residual phase-invariant emission maps to 
azimuthally-symmetric rings in a 
tomogram, so any such structures should be discounted.

The spin-folded data were not corrected for the orbital motion of the
white dwarf, since for most IPs we do not 
know the amplitude or phase of the white dwarf motion.
These motions will effectively blur the spin-folded
data, although since they typically have amplitudes (\sqiglt 80 \kmps) 
which are less than the infall motion near the white dwarf, they will not
dominate. Thus the spin-cycle tomograms
can be taken as centered on the white dwarf, but blurred by the 
orbital motion of the white dwarf. 

The third step was to compute the tomogram of the spin-varying emission,
using the Fourier-filtered back-projection recipe of Robinson \etal\
(1993). Lastly, to check the result, I re-created the 
phase-resolved spectra from the tomograms (bottom panels in Fig.~1).

For reasons of space I show the result for only one emission line
for each star. For most I show \HeIIl, since a larger fraction of it
arises in the magnetosphere than for the Balmer or \HeI\ lines, and thus the
spin cycle variations are clearest. For EX~Hya and RX\,1238--38,
which don't have strong \HeIIl, I show \Hb. See the individual papers on
each star for profiles of the lines not shown.

Before presenting the data I'll review some of the assumptions of 
Doppler tomography. The method projects onto a plane of transverse
velocity versus radial velocity, but is not sensitive to motion
out of the plane. For mapping accretion disks, which are flat to a 
good approximation, this is accounted for by a simple $\sin i$ projection
factor. However, a dipole field diverts the accreting material
out of the orbital plane, so this limitation needs to be borne in mind for
intermediate polars.  Secondly, a change of phase of the spectra 
corresponds to a rotation of the tomogram. In the following I have phased 
the data so that phase 1 coincides with the optical and/or X-ray maximum of
the spin pulse (where known). How this relates to
the orientation of the dipole is model dependent, and is discussed below.

A more serious limitation is that tomography (in both back-projection and
maximum entropy implementations) assumes that each emitting
element is equally visible at all orbital phases, ignoring occultations
and optical depth effects. We shall see that this assumption is violated
in many intermediate polars, which can lead to misleading tomograms.
If one suspects that opacity effects are causing bright regions in the 
trailed spectra one can trace along the relevant S-waves to judge which 
regions of the tomogram will be anomalously bright.

\section{Individual systems}
The individual stars are shown in Fig.~1 panels $a$ to $g$.

\subsection{AO Psc}
The \HeIIl\ line in AO~Psc shows a low velocity component, brightening and
fading over the cycle, and a higher-velocity S-wave phased with maximum
blueshift at phase 1 (which is optical and X-ray pulse maximum).  Hellier
\etal\ (1991) referred to this as a `spin-wave' to distinguish it from the
term `S-wave', which is commonly used for the specific case of the orbital
modulation caused by the impact of the accretion stream with the disk. The
spin-wave maps to the diffuse area of emission at ``3 o'clock'' in the 
tomogram.  The reconstructed data is a fair match to the observed data,
although it struggles with the low velocity component (which violates the
assumption of constant brightness over the cycle). 

The finding of maximum blueshift at phase 1 led to the `accretion
curtain' model of intermediate polars, in which optical and X-ray
maximum occurs when the upper magnetic pole points away from us, and
we predominantly see emission moving towards us onto the upper pole
(Hellier \etal\ 1987; 1991). 

Thus the interpretation of the tomogram is that it shows a velocity image of
the upper accretion curtain. We can make several remarks. Firstly,
since there is little or no emission to the left of the y-axis, we are
seeing no \HeIIl\ from the lower pole. Secondly, since the velocities are
low compared to the \sqig3000-\kmps\ white-dwarf escape velocity, even
allowing for a reduction by a factor \sqig 2 through projection onto the
line of sight, we are seeing emission from relatively far out in the 
accretion curtains, at a few white dwarf radii (see Hellier \etal\ 1991
for more discussion of this).  Thirdly, the fact that the emission
subtends \sqig 110\deg\ at the origin suggests that the accretion curtain
is extended over this angle, picking up material from the disk over a 
wide range of azimuth.

\subsection{FO~Aqr}
FO~Aqr gives a similar result to AO~Psc, with a spin-wave having
maximum blueshift near phase 1 (optical pulse maximum).  Intensity
changes in the line core are again poorly handled by the tomogram ---
the symmetry of the tomography process means that all features in the
reconstructed spectra are repeated half a cycle later, but reversed in 
velocity, resulting in spurious features in the line core.  In addition to the
line-core features, the tomogram again shows diffuse emission centered
on 3 o'clock. As with AO~Psc, the emission appears to be from only one
pole, but it subtends a greater angle than in AO~Psc, suggesting 
that the curtain is even more extended azimuthally.  Again, the low
velocities imply that the emission is from the outer parts of the
magnetosphere (particularly since FO~Aqr is a high-inclination system with a
grazing eclipse of the disk, where motions will be largely in the line
of sight). This is expected since X-ray irradiation will ionize the helium
almost completely nearer the white dwarf.

The phasing accords with results from the X-ray band.
Near phase 1, when the emission is bluest, and thus (according to
the accretion curtain model) the upper pole is 
furthest from us, the X-ray pulse
shows a narrow dip suggesting that the upper pole passes briefly
over the white dwarf limb (e.g.\ Hellier 1993b).

\subsection{BG~CMi}
BG~CMi shows a lower S/N version of the behavior seen in AO~Psc and
FO~Aqr, again with greater emission around 3 o'clock, but broken up
into separate blobs due to the greater noise.
The tomogram picks up a weak spin-wave,  again with maximum
blueshift at X-ray and optical pulse maximum. 

\subsection{PQ~Gem}
The tomogram for PQ~Gem shows clear evidence for emission from both
upper and lower accretion curtains. Indeed, one can interpret the 
tomogram showing a nearly-complete azimuthal ring of emission, 
brightening at the locations to which the upper and lower poles point
(though perhaps this is too neat to be true!).

Again, even though PQ~Gem is probably at a low inclination (Hellier
1997 estimated 30\deg), the low velocities imply that we are
seeing emission from several white dwarf radii out. 

Phase 1 is again the maximum of the blue-optical and hard-X-ray  spin pulse,
which normally locates the phase at which the poles point towards and away
from us. Note, though, that the curtains in the  tomogram appear rotated by
\sqig 40\deg\ from the x-axis. These facts suggest that the curtains are
twisted  so that material at the magnetosphere feeds predominantly onto 
field lines at an azimuth 40\deg\ ahead of the location at which it hits
the white dwarf (considered in the rotating frame) .  Encouragingly, there
is additional evidence from the X-ray lightcurves (Mason 1997) and from
polarimetry (Potter \etal\ 1997) for the same effect, with material
accreting predominantly along field lines ahead of the magnetic poles. 

One would expect the brighter of the poles to be the upper pole. 
Since the brighter pole in the tomogram is at lower-left, this
implies that the upper pole points towards us at blue-optical and 
hard-X-ray pulse maximum. This is the opposite phasing to that deduced
for the three stars above, but this phasing is also supported by the
X-ray and polarimetric results of Mason (1997) and Potter \etal\ (1997).
Perhaps since PQ Gem shows double-humped pulses in the red and in 
soft X-rays (in contrast to nearly sinusoidal pulses for the
other stars) it is not surprising that it is anomalous. See Hellier (1997),
Mason (1997) and Potter \etal\ (1997) for a fuller discussion of these
issues.

\subsection{RX\,1712--24}
RX\,1712--24 probably doesn't have an accretion disk. While the
polarisation is pulsed at the 927-s spin period, the X-ray emission is
pulsed at the 1003-s beat period (Buckley \etal\ 1997), implying the 
direct impact of the accretion stream with the magnetosphere
(e.g.\ Hellier 1991). Thus we might expect different behavior in 
RX\,1712--24.

The spectroscopy presented here is previously unpublished data obtained
on the 3.9-m AAT on the nights 1996 May 11--13 at a resolution of 
1.5 \AA\ with the RGO spectrograph. A full account will be presented
elsewhere. The phasing of the data is arbitrary since we don't yet have 
an ephemeris for the spin pulse.
The narrowness of the lines and the minimal modulation at the orbital 
period imply a low inclination for RX\,1712--24. 

The phase-resolved spectra show a central core, brightening and fading over
the cycle, together with red and blue wings brightening in phase with each
other but in anti-phase with the core.  The violation of the 
constant-brightness assumption means that the tomogram does not do a good job, as can
be seen from the reconstructed profile, where the algorithm has tried to
fill out the features in phase. The phase-resolved spectra, though, can be
interpreted by comparison with the other stars. The simultaneous blue and
red wings suggest that we are seeing emission from material falling to both
poles. Optical depth effects (or similar) prevent these from creating full
spin-waves, as seen in AO~Psc. 

\subsection{EX~Hya}
EX~Hya and RX\,1238--38 both have weak \HeIIl\ lines which show little
spin modulation, thus I show \Hb\ lines instead. Note, also,
that since the lines are much broader in these two stars, the data are
presented on a different scale. 

EX~Hya poses particular difficulties (not encountered in the other 
systems considered here) since the spin period is almost 
exactly 2/3\up{rds}\ of the orbital period, and thus orbital-cycle
variations do not smear out when folded on spin phase, but instead
repeat every 3 spin cycles.  This is the
origin of most of the structure in the line at velocities $<$\,1000 \kmps\
(see Hellier \etal\ 1987). Aside from this, the line profiles show 
broad red and blue wings at phase 1.0, fading at phase 0.5. The fact that
changes in intensity dominate over velocity variations means that the
tomogram does a poor job, showing little of interest. 

The higher velocities in EX~Hya imply that we are
seeing emission from nearer the white dwarf, $<1R_{\rm wd}$ from the
surface (see Hellier \etal\ 1987). The simultaneously-bright red and
blue wings, and the general symmetry of the line profiles, mean that we
are seeing emission from both poles. The phasing suggests that the upper
pole points away from the observer at phase 1 (optical and X-ray
pulse maximum), when the parts of the accretion curtain near the white
dwarf are readily visible. Half a cycle later, the outer regions of
the upper curtain obscure the white dwarf and the high velocity material
near it, resulting in the fainter, narrower line profile at phase 0.5. This
change in the visibility of the emitting regions prevents us seeing
full spin-waves, which would otherwise produce a red wing and a blue wing
at phase 0.5 as continuations of the blue wing and red wing (respectively)
at phase 1.

\subsection{RX\,1238--38}
RX\,1238--38 was identified as an intermediate polar by 
Buckley \etal\ (1998). It has many similarities to EX~Hya, including
similar spin and orbital periods (e.g.~Hellier, Beardmore \&\ Buckley 1998)
and similar spectroscopic behavior, including a partial eclipse of the disk.
The data presented are unpublished AAT spectra from the same run as the
RX\,1712--24 data described above.  The phasing is arbitrary, since we don't
yet have an ephemeris for the spin cycle. 
Again, intensity changes in the \Hb\ line dominate over velocity
changes and thus the tomography is not helpful,
producing a poor reconstruction of the data. The overall symmetry
of the profile, the broad wings, and the intensity changes, point to
the same explanation as that adopted for EX~Hya. However, the brighter
blue wing at phase \sqig 0.6 implies an asymmetry between the poles. 

\section{Discussion and Conclusions}
In four intermediate polars (EX~Hya, PQ~Gem, RX\,1712--24 \&\ RX\,1238--38)
we see line emission from material falling onto both magnetic poles of the
white dwarf. In three other stars (AO~Psc, FO~Aqr and BG~CMi) we see line
emission from only the upper pole. A possible explanation is that the three 
stars showing only the upper pole are all high inclination systems 
($>$\,60\deg) with grazing eclipses or X-ray dips (see, e.g., Hellier 1995).
They also show line emission only from the outer parts of the accretion 
curtains, judging from the low velocities of the spin-cycle variations. 
The outer regions of the lower accretion curtain could be hidden in a high 
inclination system, obscured by the inner edge of the accretion disk when it
points towards us and obscured by the upper accretion curtain when it
points away, explaining the lack of observed emission.

The stars showing emission from both accretion curtains are either low
inclination systems (PQ~Gem \&\ RX\,1712--24 at \sqig 30\deg) in which the
lower pole would not be obscured by the inner disk, or they are high
inclination systems below the period gap (EX~Hya \&\ RX\,1238--38). In the
latter stars we are seeing higher velocity material much closer to the white dwarf
(perhaps as a result of a lower accretion rate) so again the lower pole will
not be obscured by the disk. 

Tomography of three of the stars (AO~Psc, FO~Aqr and PQ~Gem) reveals the
azimuthal extent of line emission, showing that accretion curtains extend
over angles $>$\,100\deg\ (although note that line emission can emerge
from regions carrying only a small fraction of the accretion flow).
Further, tomography of PQ~Gem shows a twist in the accretion
curtains with material feeding onto field lines \sqig 40\deg\ 
ahead of the location at which it hits the white dwarf (see \S 3.4).

Such results show that spin-cycle tomography can be a valuable tool
for studying accretion in intermediate polars. However, the changing
optical depth through the curtain as its aspect changes 
causes large emission-line flux changes over the spin cycle. Since this
is not taken into account, tomography can give misleading results
in some systems, so needs to be interpreted with caution. 

\acknowledgments
I thank collaborators including Keith Mason, Mark Cropper, Rob Robinson,
Dave Buckley and Alasdair Allan for their involvement in obtaining some 
of the spectroscopy presented here. I thank the referee for prompting 
several clarifications of the manuscript.

\figcaption{The line profiles and tomograms for either \HeIIl\ or \Hb\ for
each star. Darker colouring implies greater intensity.  The top panels are
the line profiles folded on the spin cycle. In all cases the data were
normalised to the continuum so the plots show quasi-equivalent widths
(note that this process is not the cause of the line intensity changes seen,
since the lines are generally pulsed to a greater extent than the
continuum). The
panels second from top show the same data after subtraction of the
phase-invariant profile. The third panels show the Doppler tomograms,
computed by Fourier-filtered back projection. The rotation of the tomograms
in this paper conforms to the standard presentation of orbital-cycle
tomograms, where emission in the 12 o'clock location produces an S-wave with
blue-to-red crossing at phase 1. The lowest panels show the line profiles
reconstructed from the tomograms.}

\newpage

\begin{table}[t]\caption{The intermediate polars discussed in this paper}
\begin{center}\begin{tabular}{l@{}ccc}\tableline
Star\rule{0mm}{4mm} & Spin period & Orbit & Original paper \\ 
   &  (secs)  &  (hours) \\ [0.5mm] \tableline
AO~Psc\rule{0mm}{4mm} &  805 & 3.59 & Hellier \etal\ 1991 \\
FO~Aqr       &  1254 & 4.85 & Hellier \etal\ 1990 \\
BG~CMi       &  913 & 3.23 & Hellier 1997 \\
PQ~Gem       &  833 & 5.19 & Hellier 1997 \\
RX\,1712--24 &  927 & 3.42 & Unpublished \\
EX~Hya       &  4022 & 1.64 & Hellier \etal\ 1987 \\ 
RX\,1238--38 &  2147 & 1.42 & Unpublished \\ [0.5mm] \tableline 
\end{tabular}\end{center}\end{table}
\rule{0mm}{5mm}\rule{0mm}{5mm}
\rule{0mm}{5mm}

\newpage
\vspace*{21cm}

\includegraphics{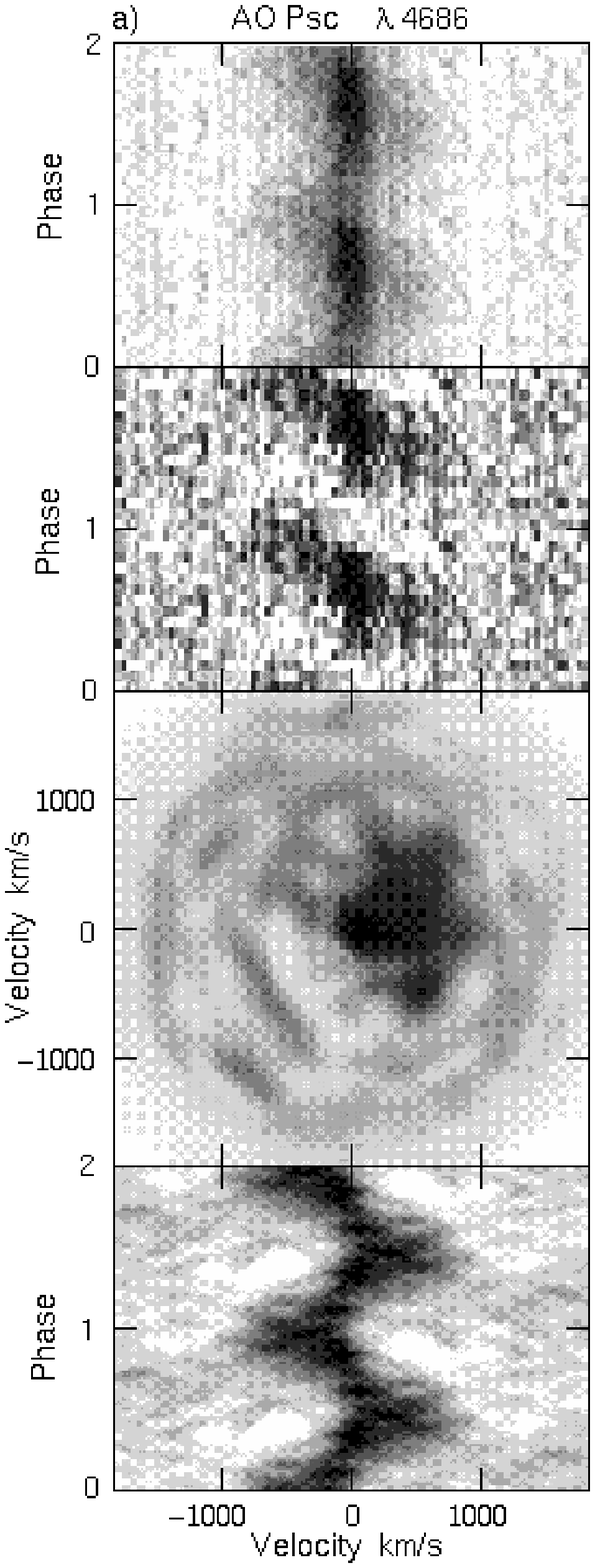}

\newpage
\vspace*{21cm}

\includegraphics{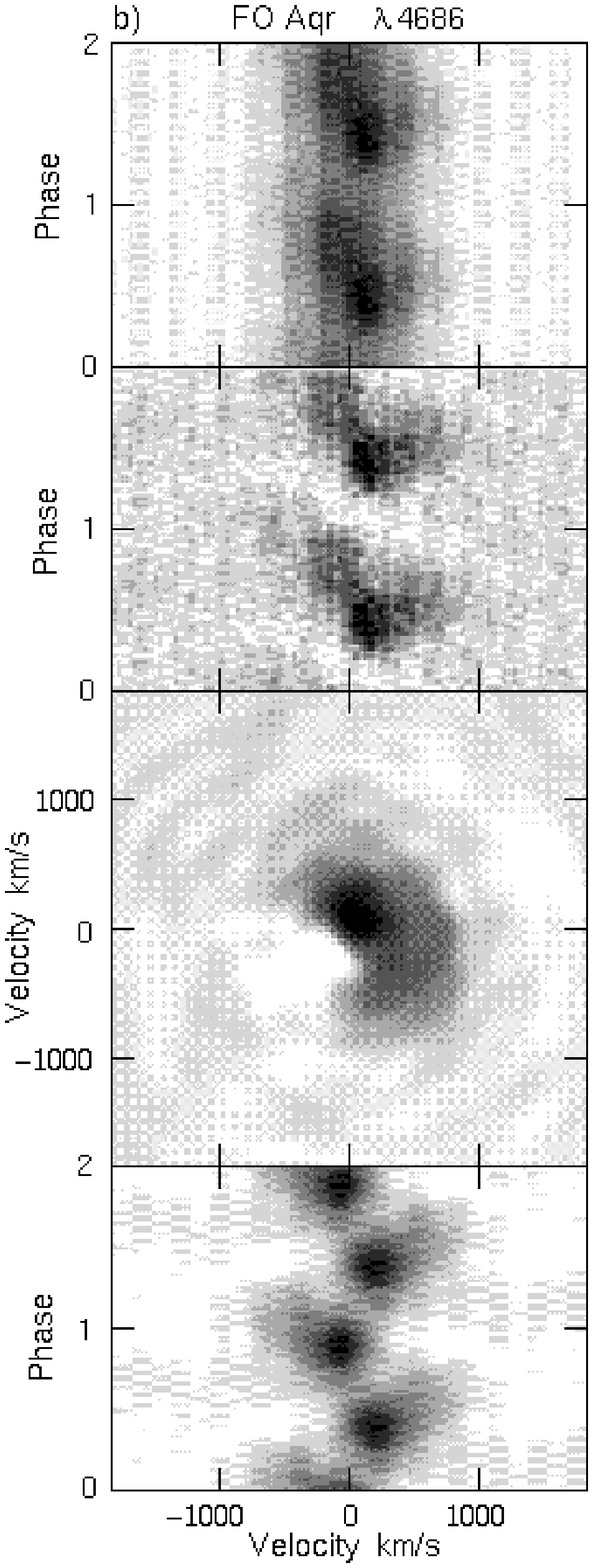}

\newpage
\vspace*{21cm}

\includegraphics{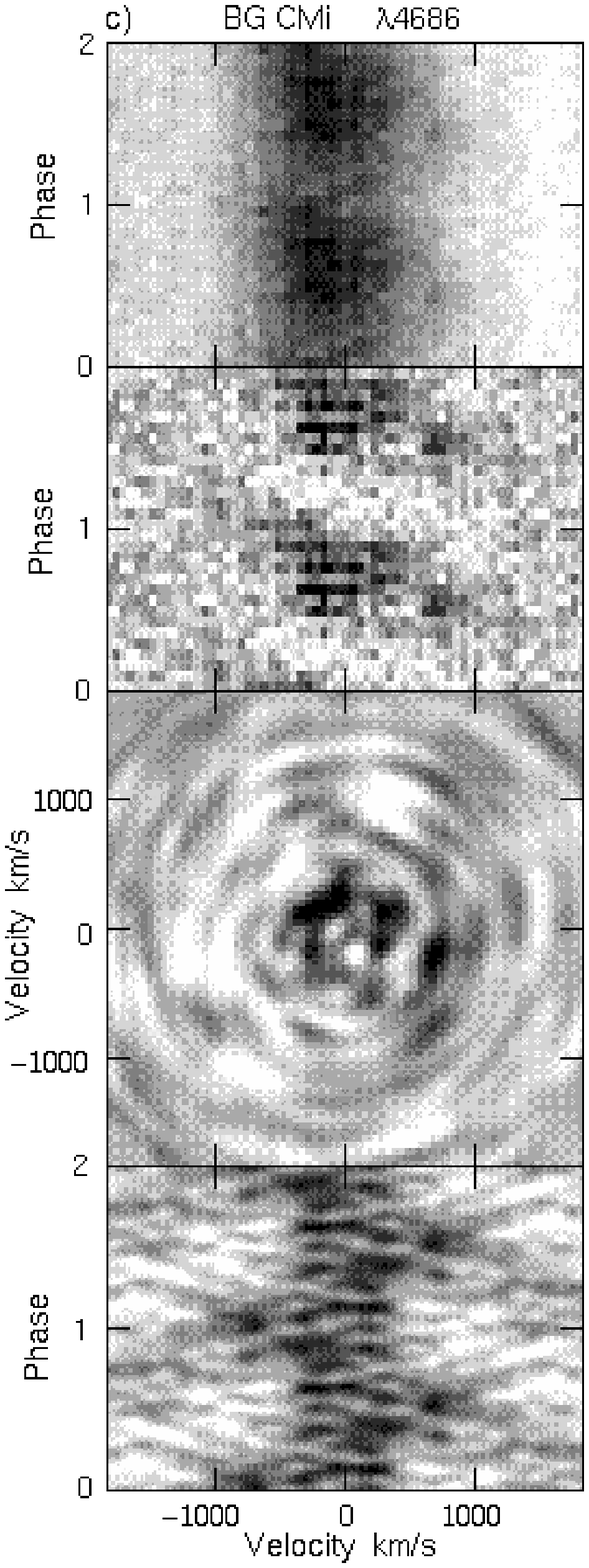}

\newpage
\vspace*{21cm}

\includegraphics{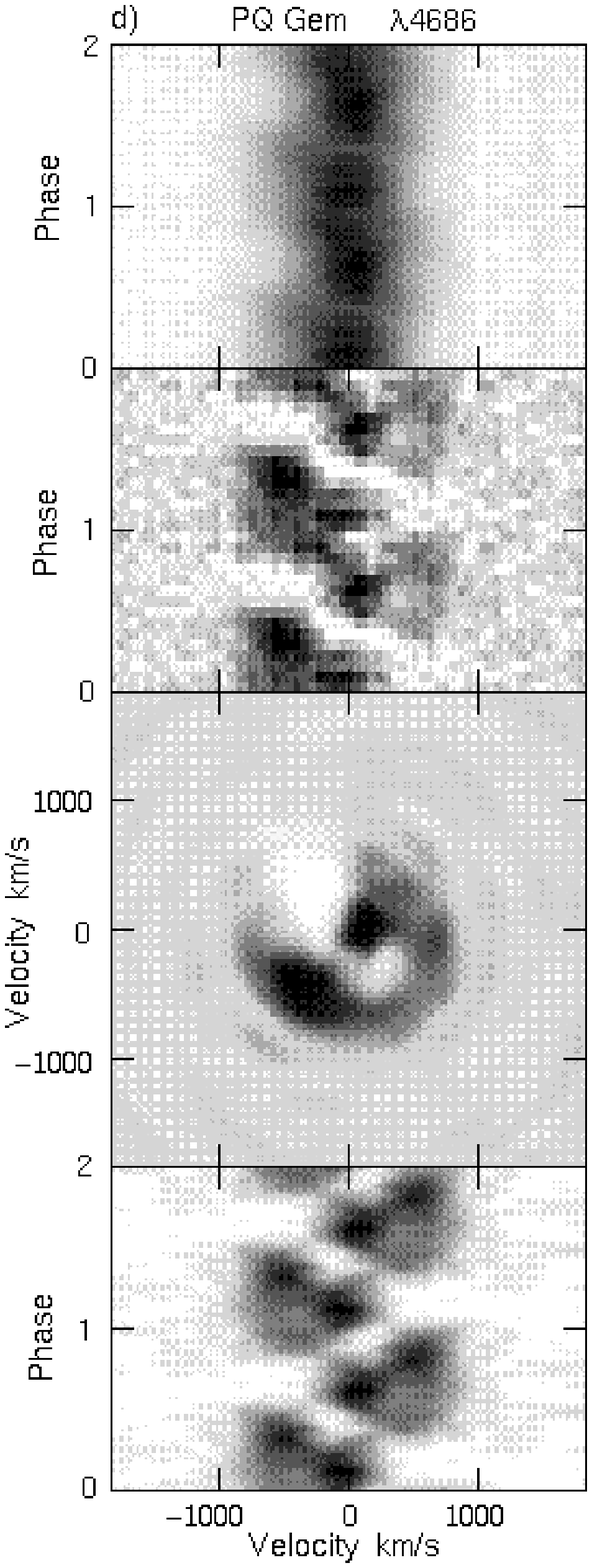}

\newpage
\vspace*{21cm}

\includegraphics{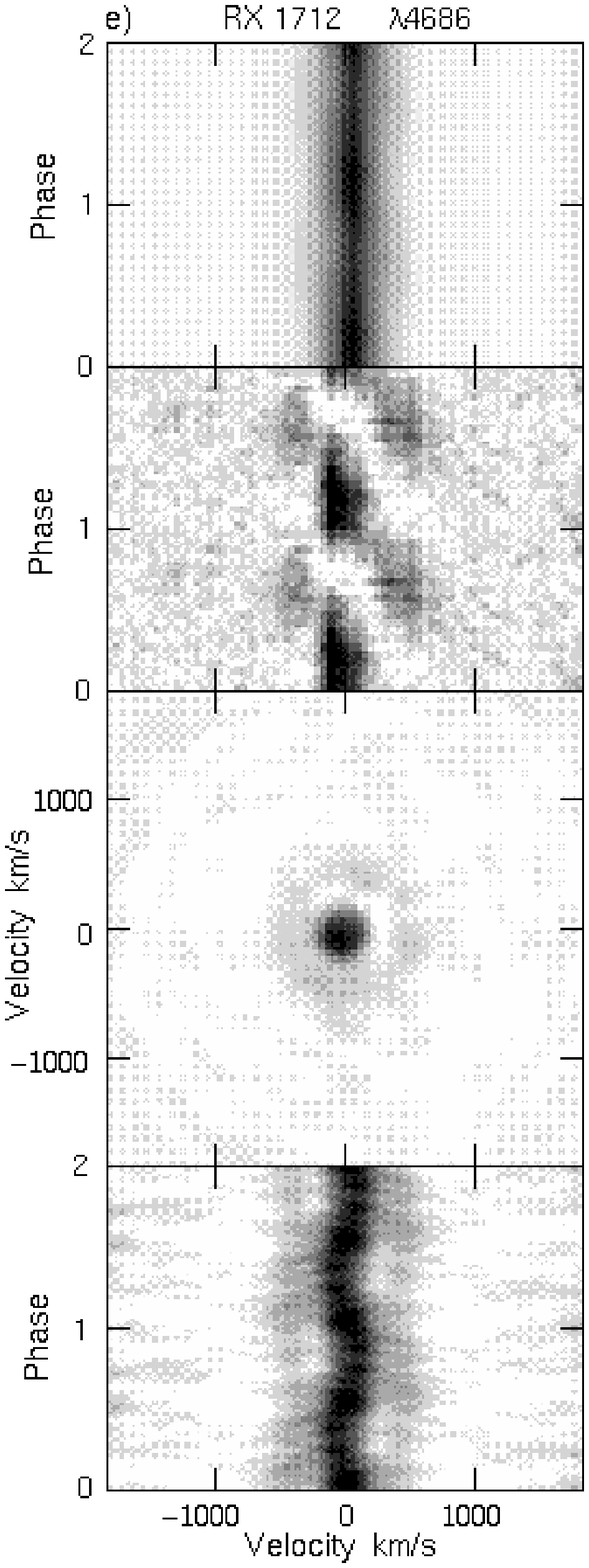}

\newpage
\vspace*{21cm}

\includegraphics{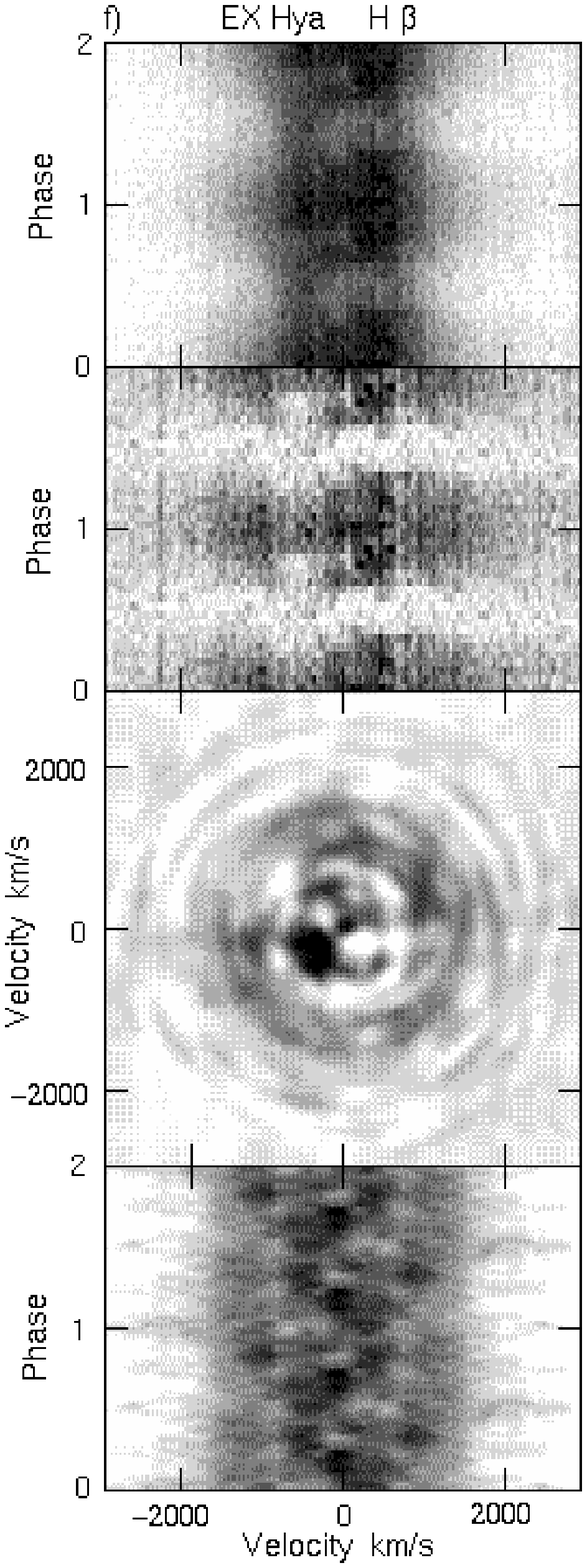}

\newpage
\vspace*{21cm}

\includegraphics{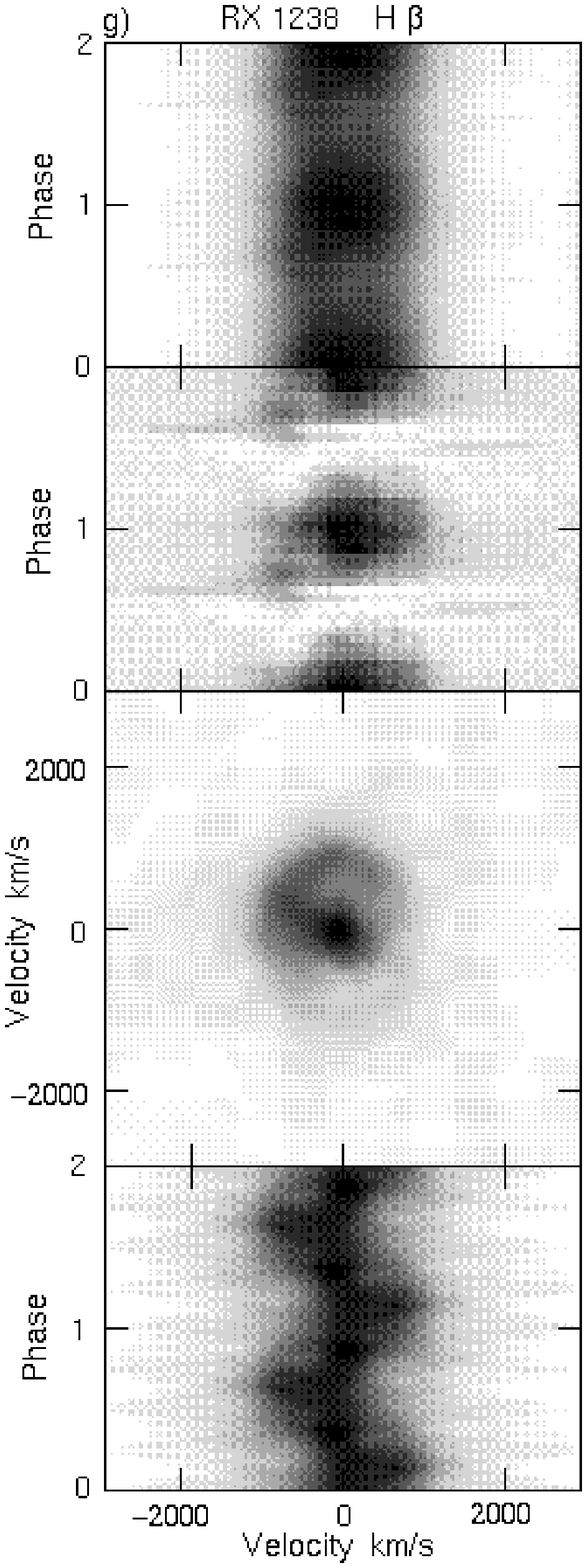}


\begin{thebibliography}{}
\bibitem[]{}Buckley D.\,A.\,H., Haberl F., Motch C., Pollard K., 
Schwarzenberg-Czerny A., \&\ Sekiguchi K., 1997, MNRAS, 287, 117
\bibitem[]{}Buckley D.\,A.\,H., Cropper M., Ramsay G., Wickramasinghe D.\,T.
1998, MNRAS, in press
\bibitem[]{}Hellier C. 1991, MNRAS, 251, 693
\bibitem[]{}Hellier C. 1993a, MNRAS, 264, 132
\bibitem[]{}Hellier C. 1993b, MNRAS, 265, L35
\bibitem[]{}Hellier C. 1995, in Cape workshop on magnetic cataclysmic 
variables, eds, D.\,A.\,H.~Buckley, \&\ B.~Warner, ASP Conf series, 85, p185
\bibitem[]{}Hellier C. 1997, MNRAS, 288, 817
\bibitem[]{}Hellier C., Beardmore A.\,P., \&\ Buckley D.\,A.\,H. 1998, 
   MNRAS, 299, 851
\bibitem[]{}Hellier C., Cropper M., \&\ Mason K.\,O. 1991, MNRAS, 248, 233
\bibitem[]{}Hellier C., Mason K.\,O., \&\ Cropper M. 1990, MNRAS, 242, 250
\bibitem[]{}Hellier C., Mason K. O., Rosen S. R., \&\ Cordova F. A. 1987, MNRAS,
   228, 463
\bibitem[]{}Kaitchuck R.\,H., Schlegel E.\,M., Honeycutt R.\,H., Horne K.,
   Marsh T.\,R., White J.\,C., \&\ Mansperger C.\,S. 1994, ApJS, 93, 519
\bibitem[]{}Marsh T.\,R., \&\ Horne K. 1988, MNRAS, 235, 269
\bibitem[]{}Mason K.\,O. 1997, MNRAS, 285, 493
\bibitem[]{}Patterson J. 1994, PASP, 106, 209
\bibitem[]{}Potter S.\,B., Cropper M., Mason K., Hough J.\,H., \&\ 
   Bailey J.\,A. 1997, MNRAS, 285, 82
\bibitem[]{}Robinson E.\,L., Marsh T.\,R., \&\ Smak J.\,I. 1993, in
   `Accretion disks in compact stellar systems', ed, J.\,C.\,Wheeler 
    (Singapore: World Scientific), 75
\bibitem[]{}Schwope A.\,D., Mantel K.-H., \&\ Horne K. 1997, A\&A, 319, 894 
\bibitem[]{}Still M.\,D., Duck S.\,R., Marsh T.\,R., 1998, MNRAS, 299, 759
\bibitem[]{}Vogt S.\,S., \&\ Penrod G.\,D. 1983, PASP, 95, 565
\end{thebibliography}
\end{document}